\documentclass[sigconf]{acmart}

\usepackage{colortbl}

\usepackage{tabularx} 
\usepackage{booktabs}
\usepackage{caption}
\usepackage{footnote}

\usepackage{colortbl}
\usepackage{enumitem}

\usepackage{caption}
\usepackage{subcaption}

\usepackage{multirow}  

\usepackage{tabularx}

\usepackage[multiple]{footmisc}

\usepackage{array}

\newcommand{\tabitem}{~~\llap{\textbullet}~~} 

\AtBeginDocument{%
  \providecommand\BibTeX{{%
    \normalfont B\kern-0.5em{\scshape i\kern-0.25em b}\kern-0.8em\TeX}}}

\copyrightyear{2024}
\acmYear{2024}
\setcopyright{rightsretained}
\acmConference[Websci '24]{ACM Web Science Conference}{May 21--24,
2024}{Stuttgart, Germany}
\acmBooktitle{ACM Web Science Conference (Websci '24), May 21--24, 2024,
Stuttgart, Germany}\acmDOI{10.1145/3614419.3644004}
\acmISBN{979-8-4007-0334-8/24/05}

\begin{document}

\title{Corrective or Backfire: Characterizing and Predicting User Response to Social Correction}

\author{Bing He, Yingchen Ma, Mustaque Ahamad, Srijan Kumar}
\affiliation{%
  \institution{Georgia Institute of Technology}
  \city{Atlanta}
  \state{Georgia}
  \country{USA}
}
\email{bhe46@gatech.edu, yma473@gatech.edu, mustaq@cc.gatech.edu, srijan@gatech.edu}

\begin{abstract}

Online misinformation poses a global risk with harmful implications for society. 
Ordinary social media users are known to actively \textit{reply} to misinformation posts with counter-misinformation messages, which is shown to be effective in containing the spread of misinformation. 
Such a practice is defined as ``\textit{social correction}''. 
Nevertheless, it remains unknown how users respond to social correction in real-world scenarios, especially, will it have a corrective or backfire effect on users. 
Investigating this research question is pivotal for developing and refining strategies that maximize the efficacy of social correction initiatives.

To fill this gap, we conduct an in-depth study to characterize and predict the user response to social correction in a data-driven manner through the lens of X (Formerly Twitter), where the user response is instantiated as the reply that is written toward a counter-misinformation message.
Particularly, we first create a novel dataset with $55,549$ triples of misinformation tweets, counter-misinformation replies, and responses to counter-misinformation replies, and then curate a taxonomy to illustrate different kinds of user responses. 
Next, fine-grained statistical analysis of reply linguistic and engagement features as well as repliers' user attributes is conducted to illustrate the characteristics that are significant in determining whether a reply will have a corrective or backfire effect. 
Finally, we build a user response prediction model to identify whether a social correction will be corrective, neutral, or have a backfire effect, which achieves a promising F1 score of 0.816. 
Our work enables stakeholders to monitor and predict user responses effectively, thus guiding the use of social correction to maximize their corrective impact and minimize backfire effects.
The code and data is accessible on \url{https://github.com/claws-lab/response-to-social-correction}.

\end{abstract}



\settopmatter{printfolios=true} 

\maketitle

\begin{figure*}[!thbp]
    \centering
    \includegraphics[width=1\textwidth]{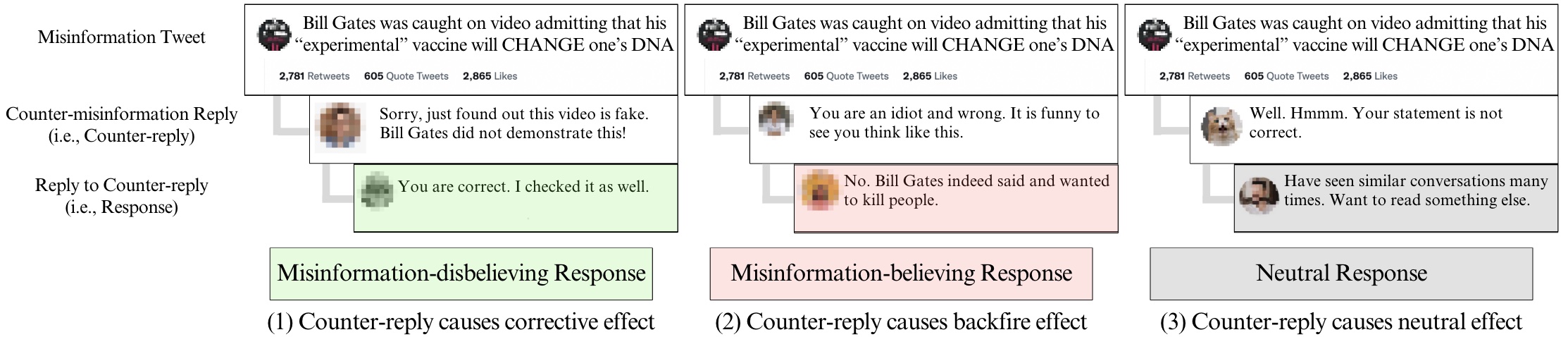}
    \caption{Examples of user responses to social correction. Here, the social correction is the counter-misinformation \textit{reply} posted by ordinary users (the second row), and the user response is the \textit{reply} to the counter-misinformation reply (the third row). 
    } 
    \label{fig:examples}
\end{figure*}
\section{Introduction}

Online misinformation undermines public health by diminishing trust in vaccines and health policies~\cite{pierri2022online,ball2020epic,lazer2018science}, and has been linked to reduced COVID-19 vaccine uptake~\cite{ma2023characterizing}. 
Its impact also extends to inciting violence~\cite{starbird2014rumors,arif2018acting}, and negatively affecting well-being~\cite{verma2022examining}. 
This situation is exacerbated because misinformation typically spreads more rapidly and widely than factual information on online social media platforms~\cite{vosoughi2018spread,lazer2018science}, making it imperative to curb the spread of misinformation~\cite{Micallef2020, lewandowsky2012misinformation, Goindani2020, Budak2011, Zhu2021, Litou2017, Wang2020}.

To combat misinformation, professional fact-checkers and journalists provide valuable objective fact-checks to debunk misinformation~\cite{vo2019learning}. 
However, their engagement with users is limited~\cite{Micallef2020}. In contrast, ordinary social media users play a proactive role in combating misinformation through their active engagement including their replies, comments, and posts that counter misinformation posted by others~\cite{Micallef2020,Tully2020,vo2018rise,zhou2022fake,Bode2018,starbird2014rumors}. It finally complements the efforts of professionals~\cite{Allen2021,kim2018leveraging, he2023survey}, even accounting for 96\% of online counter-misinformation messages~\cite{Micallef2020}. 

Significantly, recent studies underscore the ``\textit{social correction}''~\cite{ma2023characterizing, miyazaki2023fake} - the practice where ordinary users combat misinformation claims in a \textit{conversational} manner by their direct counter misinformation \textit{replies} to misinformation posts - which has shown to be as effective as professional correction, 
curbing misinformation spread across diverse topics, platforms, and demographics~\cite{ma2023characterizing,seo2022if,friggeri2014rumor,colliander2019fake,Wijenayake2021,bode2020right,vraga2018not,vraga2021addressing,bode2015in,Bode2018,vraga2020correction,vraga2021addressing,vraga2021assessing,vraga2021effects}. One example of social correction is shown in Figure~\ref{fig:examples}.

Nevertheless, little is known about the real-world user response toward social correction. 
Understanding such responses is beneficial because 
i) They serve as a critical signal to indicate the impact of social correction in real-world scenarios. If some social corrections are revealed to have corrective effects (e.g., users disbelieve in misinformation)~\cite{chan2017debunking}, then additional participants can be encouraged to provide reinforcements;
ii) Instead, If certain social corrections are found to increase users' beliefs in misinformation (e.g., backfire)~\cite{swire2022backfire}, targeted efforts can be directed toward improving them. Such instances can be escalated and prioritized for interventions by professionals or social media platforms; 
iii) Responses can also indicate whether users are entrenched in (counter-)misinformation echo chambers~\cite{del2015echo}, where their beliefs are reinforced by similar viewpoints, or if there is a cross-pollination of ideas. This contributes to understanding polarization around certain topics.

Despite its advantages, characterizing and predicting social correction is challenging because of multiple reasons.
First, current research predominantly utilizes simulated experiments or user studies~\cite{Borah2021, Kirchner2020,  Tully2020}, approaches that may not accurately mirror real-world scenarios.
Second, relevant research works and datasets~\cite{wang2022factors, zhang2022investigation} do not consist of conversational-style narratives with tripled misinformation posts, counter-replies, and responses, as shown in Figure~\ref{fig:examples}. 
The analysis of these conversations can reveal the complex interactions among misinformation spreaders, those who counter-reply, and the responders to these counter-replies. 
It plays a vital role in demonstrating the organic processes of both correcting and exacerbating misinformation. 
Third, related research works do not conduct fine-grained investigation of user responses. The traditional four-class stance~\cite{wang2022factors} or two-class sentiment~\cite{zhang2022investigation} categorization of user responses only provides a shallow classification of user responses. A more comprehensive taxonomy of user actions behind their responses is needed to provide a better understanding of how users respond differently to social correction.

To address these challenges, we first curate conversational-style user response datasets to social correction on Twitter\footnote{Twitter was renamed as ``X'' in July 2023. We continue to refer to the platform as ``Twitter'' for illustration.} and create the taxonomy of these user responses. 
Additionally, we conduct a statistical analysis of linguistic-, engagement-, and poster-level characteristics of counter-replies to examine user responses to social correction at a fine-grained level. 
Finally, we create a prediction model to forecast whether a counter-reply will have a corrective, backfire, or neutral effect on users. 
In sum, the contributions of the paper are summarized as follows:
\begin{itemize}
    \item We curate a novel large-scale dataset that contains 1,523,849 misinformation tweets, 254,779 counter-misinformation replies, and 55,549 responses, along with a hand-annotated dataset of misinformation tweets, counter-replies, and counter-replies. Concurrently, we build a taxonomy of user responses to demonstrate different kinds of responses to social correction.
    \item We perform a fine-grained analysis of the linguistic, engagement, and poster-level characteristics of counter-replies that have a corrective or backfire effect. Our analysis reveals several salient features of counter-replies that are more common in corrective replies (e.g., politeness (4.678\%), evidence (8.137\%), and positiveness (5.409\%)) than in backfire replies. 
    \item We create a user response prediction model to identify whether a counter-reply will be a corrective, backfire, or neutral reply. The model achieves a promising predictive performance with an F1 score of $0.816$.
\end{itemize}

The code and data is accessible on \url{https://github.com/claws-lab/response-to-social-correction}.
\section{Related Works}

\subsection{Social Correction on Social Media Platforms}

Misinformation causes harms to the society~\cite{pierri2022online,ball2020epic,lazer2018science, ma2023characterizing, starbird2014rumors,arif2018acting}.
Social correction by ordinary users (i.e., the crowds) is a crucial approach to combating misinformation~\cite{ma2023characterizing, miyazaki2023fake} due to its accessibility and scalability to complement professional fact-checkers~\cite{Micallef2020}. 
It has also been shown to be effective by conducted 
interviews~\cite{Borah2021,Kirchner2020,Tully2020}, surveys~\cite{Veeriah2021,Kirchner2020}, and in-lab experiments~\cite{Tully2020} in
curbing misinformation spread~\cite{friggeri2014rumor,colliander2019fake,Wijenayake2021} across topics~\cite{bode2020right,vraga2018not,vraga2021addressing,bode2015in,Bode2018,vraga2020correction}, platforms, and demographics~\cite{vraga2021addressing,vraga2021assessing,vraga2021effects}. 
To facilitate broad social correction among users, Twitter launched the Community Note system~\footnote{Community Note is previously known as Birdwatch.}~\cite{prollochs2022community}, where users can report and flag potential misinformation tweets. 
These works and practices demonstrate that social correction works. 
However, user response to social correction is less studied in the existing research works despite its benefits in understanding the efficacy of social correction. More critically, current works rely on simulated experiments and user studies~\cite{Tully2020} and do not study datasets that represent real-world scenarios. 
To bridge this research gap, we conduct a large-scale data-driven analysis. We specifically focus on examining user replies to counter-misinformation replies on Twitter, thereby providing insights grounded in real-world response behavior.

\subsection{User Response to Misinformation Correction}

To correct misinformation, ordinary users can publish standalone counter-misinformation posts on social media platforms~\cite{Micallef2020}.
User responses to this kind of correction have been investigated~\cite{wang2022factors, zhang2022investigation, wang2021evaluating}. 
For instance, \citet{wang2022factors} analyze the comments on fake news rebuttal posts through the expressed stance in them. They find that information readability and argument quality improve the acceptance of misinformation rebuttal. They also uncover that citing evidence helps~\cite{wang2021evaluating}. 
\citet{zhang2022investigation} similarly investigate the sentiment in comments that respond to fact-checking posts. But, all these posts are from official fact-checking organization accounts~\cite{zhang2022investigation}, which is different from our setting of ordinary users. 
Additionally, none of these corrections occur in a conversational manner like our focus of social correction that has more engagement and visibility between misinformation spreaders and those who counter-reply~\cite{ma2023characterizing}. 
These existing conventional four-class stance~\cite{wang2022factors} or two-class sentiment~\cite{zhang2022investigation} studies only provide coarse-grained analysis of user responses.

Some researchers examine another type of misinformation correction - the warning labels posted around the misinformation posts. 
For example, ~\citet{chuai2023roll} focus on labels as well as the associated fact-check text provided voluntarily by users within Twitter's Community Note system~\cite{prollochs2022community}. Different from our response analysis, they only focus on the volume of retweets and likes of the fact-checked tweet. However, retweets and likes are all non-negative signals and are unable to comprehensively capture the user response, especially, the negative responses.
In addition, users provide inputs within the Community Note system only, which is restricted (e.g., users cannot write responses to the fact-checking text and labels on the Twitter platform) and does not reflect the larger dynamics of information flow on Twitter. 

\subsection{Backfire and Corrective Effects of Misinformation Correction}

When misinformation is debunked, it may have a backfire effect,  i.e., users viewing the counter-misinformation post or misinformation spreaders potentially increase their belief in the misinformation due to observing the correction. This has been debated for a long time~\cite{lewandowsky2012misinformation, nyhan2010corrections}. 
Even if some researchers find the backfire effect among particular groups~\cite{schmid2022benefits} and within certain time frames~\cite{peter2016debunking}, many studies have failed to replicate the backfire effect~\cite{swire2020searching, wang2018rumor}. 
On the other hand, corrective effects, i.e., the audience or the misinformation spreaders instead decrease their belief in misinformation after viewing the counter-misinformation, have been identified by existing research works~\cite{chan2017debunking,walter2021evaluating,walter2018unring,porter2021global, seo2022if, friggeri2014rumor}. 
Nevertheless, the existing studies of backfire and corrective effects usually leverage simulated experiments to examine their hypothesis about backfire and corrective effects while neglecting real-world scenarios, especially the situations where misinformation is corrected by real-world ordinary users rather than professionals or bot accounts. 
To fill this gap, we examine these effects through real-world user replies to counter-misinformation posts in a data-driven manner.
Since this user response information can indicate the effects of certain textual properties in counter-replies, our work can lead to a better understanding of the impacts of social correction behavior, especially, comprehending the counter-replies that are corrective or backfire. 
\section{Dataset} 

\subsection{Definition}

\subsubsection{\textbf{Misinformation Tweet}}\label{sec:misinfo_def}

We deploy a broad definition of misinformation which includes inaccuracies, falsehoods, rumors, or misleading leaps of logic~\cite{wu2019misinformation}. Based on the existing work~\cite{hayawi2022anti, ma2023characterizing}, we focus on misinformation related to the COVID-19 vaccine due to its broad impact around the world during the COVID-19 pandemic. 
Particularly, the misinformative claims include ``the vaccine changes genes'', ``the vaccine leads to infertility'', ``the vaccine is created by Bill Gates to kill people'', and ``the vaccine consists of microchips to control people''; 
these misinformation topics are widely studied by existing research works due to their popularity~\cite{hayawi2022anti, abbasi2022widespread, he2023reinforcement}.
The misinformation tweet is represented as $m$.

\subsubsection{\textbf{Counter-misinformation Reply (i.e., Counter-reply)}}\label{sec:counter_reply_def}

Inspired by existing research works on social correction~\cite{ma2023characterizing, miyazaki2023fake}, a direct reply to a misinformation tweet $m$ is considered as a counter-misinformation reply (i.e., counter-reply as shown in Figure~\ref{fig:examples} and denoted as $c$), if it attempts to counter the misinformation tweet. 
Particularly, building on existing research works that identify and analyze the text that is countering, debunking, disbelieving, or disagreeing with misinformation~\cite{Micallef2020, jiang2020modeling, hossain2020covidlies, ma2023characterizing, miyazaki2023fake}, a counter-reply is a reply that explicitly or implicitly refutes the misinformation post (``the tweet is wrong. it is misinformation''), targets the tweet poster (``you are born to speak nothing but lies''), or highlights the falsehood (``the COVID-19 vaccine does not change DNA''). 

\subsubsection{\textbf{Reply to Counter-reply (i.e., Response)}}\label{sec:response_def}

On Twitter, users can respond to a counter-reply via a direct reply to it, as shown in Figure~\ref{fig:examples}.
These responses denote the responder's 
stance toward misinformation, serving as a crucial signal to study the impact of counter-reply. 
Following existing work on similar stance analysis~\cite{wang2021evaluating, wang2022factors,jiang2020modeling}, we can group responses into three categories, as shown in Figure~\ref{fig:examples}: 
\begin{itemize}
    \item Misinformation-disbelieving responses:
    responses disbelieve in misinformation or believe in counter-misinformation (e.g., ``You are correct. I checked it as well.'');  
    \item Misinformation-believing responses: 
    responses believe in misinformation or disbelieve in counter-misinformation (e.g., ``No, Bill Gates indeed said and wanted to kill people''); 
    \item  
    Neutral responses:
    Responses neither believe nor disbelieve in misinformation, lacking sufficient information for judgment (e.g., `` Have seen similar conversation many times. Want to read something else.''). %
\end{itemize}

\subsection{Task Objective}\label{sec:task_objective}

Given the set $\mathcal{M}$ of misinformation posts regarding the COVID-19 vaccine, each misinformation post $m \in \mathcal{M}$ has a set of $n$ counter-replies $c = [c_1, c_2, ..., c_n]$ posted in direct reply to $m$. 
Our goal is to build a classifier $\mathcal{F}$ such that it can output a label $\mathcal{F}(c_i), i \in \{1, 2, ..., n\}$, which indicates whether the counter-reply will have a corrective, backfire, or neutral effect, i.e., the counter-reply will have at least one misinformation-disbelieving response but no misinformation-believing responses (corrective), at least one misinformation believing response but no misinformation-disbelieving response (backfire), or only neutral responses (neutral)?
\footnote{Note that we do not emphasize the case where a counter-reply has both misinformation-believing and misinformation-disbelieving responses, which can be worth exploring in future studies, because (1) it has the lowest volume, accounting for only $0.93\%$ of all $254,779$ counter-replies in our dataset, as shown in Section~\ref{sec:counter_reply_group}; and (2) similar existing research works also do not emphasize this kind of dual labels~\cite{he2021racism}.}

\subsection{Dataset Curation}

\subsubsection{\textbf{Misinformation Tweet Collection and Classification}}

In our study, we followed an existing approach~\cite{ma2023characterizing} and used the Anti-Vax dataset from~\citet{hayawi2022anti}, containing around 15.4 million English tweets about COVID-19 vaccines, collected between December 1, 2020, and July 31, 2021. These tweets, which exclude retweets, replies, and quotes, were filtered to include key vaccine-related terms (e.g., ``vaccine'', ``pfizer'', and ``moderna''). From the original set, 14,123,209 tweets are retrievable via the Twitter API while the remaining 1.3 million tweets are unavailable due to deletion by users or Twitter.

To identify misinformation tweets, we followed the definition outlined in Section~\ref{sec:misinfo_def} and the current approach by ~\citet{hayawi2022anti}. Initially, $13,432$ annotated tweets (4,836 misinformation, and 8,596 non-misinformation) were extracted from ~\citet{hayawi2022anti}. Using these tweets, we trained a BERT-based text classifier~\cite{devlin2019bert}, achieving precision, recall, and F-1 score of $0.972$, $0.979$, and $0.975$, respectively, denoting a satisfactory performance for the misinformation classification task.

Applying this classifier to the full dataset, we identified $1,523,849$ misinformation and $12,599,360$ non-misinformation tweets. However, since we focus on replies to misinformation tweets and responses to these replies, we only keep misinformation tweets that contain this information, resulting in $44,557$ misinformation tweets.

\subsubsection{
\textbf{Counter-reply Collection and Classification}}

For each misinformation tweet, we use the Twitter API to crawl all direct replies to the original tweet. 
In total, we collect a total of $707,529$ replies to the $44,557$ tweets. One misinformation tweet has an average of approximately $16$ replies. 

To identify counter-replies, we follow the definition of counter-reply in Section~\ref{sec:counter_reply_def} and build on existing works of counter-reply classification~\cite{ma2023characterizing, he2023reinforcement}. Particularly, we first crawl a combined $2,479$ annotated replies ($1,425$ counter-replies, and $1,054$ non-counter-replies) from ~\cite{ma2023characterizing, he2023reinforcement}. 
Next, we train a RoBERTa-based lower-case counter-reply classifier~\cite{ma2023characterizing} attaining precision, recall, and F1 score of $0.801$, $0.913$, and $0.858$, respectively, which is sufficient for 
counter-reply classification. Finally, we identify $254,855$ replies as counter-replies, and the remaining as non-counter-replies.

\paragraph{Counter-reply Poster Attribute Collection}
For each counter-reply, we also collect information of the user who posted the counter-misinformation reply, which contains the date and time of account creation, the number of tweets posted, account verification, follower count, and following count. In total, information for $251,017$ unique users was retrieved.

\subsubsection{
\textbf{Response Collection and Classification}} 
For each counter-reply, we use the Twitter API to crawl all direct replies to counter-replies. In total, we collected a total of $55,549$ replies to $34,765$ counter-replies that have responses.
Because it is labor-intensive to manually annotate all $55,549$ responses, we instead aim to train a text-based classifier for annotation. 
Following the existing works of building tweet text classifiers~\cite{jiang2020modeling}, we first annotate responses and then train the classifier.
Particularly, two students annotated $601$ randomly selected responses into ``misinformation-believing'', ``misinformation-disbelieving'', and ``neutral'' as per the definition in Section~\ref{sec:response_def}. Such annotation results in an inter-rater agreement score of $0.7970$. 
After two students discuss the data points that are labeled differently and reach a consensus, 
we finally have $213$ misinformation-disbelieving responses, $218$  misinformation-believing responses, and $170$ neutral responses. 
We then fine-tune a RoBERTa-based classifier, which unfortunately has an under-satisfactory performance in precision, recall, and F1 score of $0.545$, $0.526$, and $0.511$. 
The potential reason is that one data point consists of three entries (i.e., misinformation tweet, counter-reply, and response) and there are complex inter-relationships between them. This requires a profound understanding of one data point, thus making the RoBERTa-based classification task extremely challenging. 

\begin{figure}[!tbp]
    \centering
    \includegraphics[width=0.8\columnwidth]{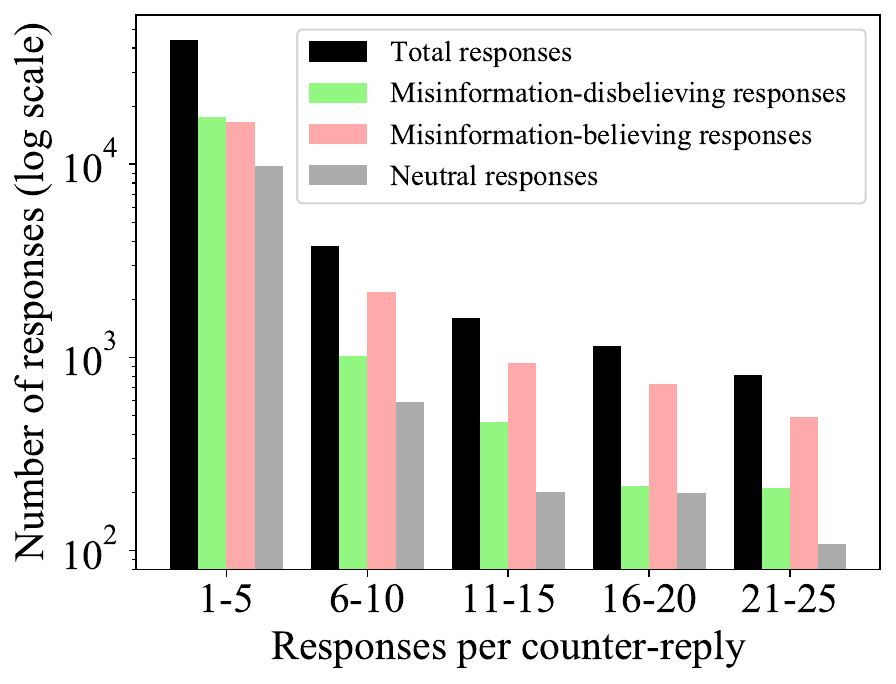}
    \caption{Distributions of the total number of responses (black), number of misinformation-disbelieving responses (green), number of misinformation-believing responses (red), and number of neutral responses (gray) per counter-reply, each presented on a log scale.}
    \label{fig:response_count}
\end{figure}

On the other hand, Large Language Models have progressed and shown the potential of accurately annotating text due to their human-level understanding of text~\cite{gilardi2023chatgpt, ziems2023can}, especially the GPT-4 model~\cite{mao2023gpteval}. 
Building on the existing research works regarding ChatGPT-based text annotation in computation social science domains~\cite{ziems2023can}, we adopt the well-performed few-shot in-context-learning diagram for GPT-4 annotation~\cite{gilardi2023chatgpt, ziems2023can}. 
First, to justify the capability of GPT-4 in annotating responses, we randomly sample four annotated data points in each category as the guidance in our carefully crafted prompt, which is presented in Figure~\ref{fig:gpt4_prompts}.
\begin{figure}[!tbp]
\parbox{\linewidth}
{
\parbox{0.93\linewidth}{
\fbox{
\parbox{\linewidth}
{
\textbf{\textit{System:}} Assume you can help people to label the reply-to-reply text. Particularly, in a JSON object, you will have "tweet", "reply", and "reply-to-reply" information where "tweet" is misinformation or false information, "reply" is countering, correcting, or debunking "tweet", and "reply-to-reply" is replying towards "reply". After understanding the content in the JSON object, you provide "label" for "reply-to-reply" where 
"-1" indicates the "reply-to-reply" disbelieves "reply" or believes in "tweet"; 
"0" means "reply-to-reply" does not  believe or disbelieve "reply", lacking sufficient information for judgment; 
"1" means "reply-to-reply" believes "reply" or disbelieves "tweet".

\textbf{\textit{User:}} 
\{
"tweet": "Bill Gates was caught on video admitting that his experimental vaccine will CHANGE one's DNA", 
"reply": "Sorry, just found out this video is fake. Bill Gates did not demonstrate this!", 
"reply-to-reply": "You are correct. I checked it as well"
\}

\textbf{\textit{GPT-4:}} \{ {"label": "1"} \}

\textbf{\textit{User:}} \{ {...} \}

\textbf{\textit{GPT-4:}} \{ {...} \}
}
}
}
}

\caption{Illustration of prompts used in GPT-4 annotation.\label{fig:gpt4_prompts}}
\end{figure}
Then, we use this prompt to label our remaining annotated responses using a suggested moderate temperature of $0.5$ in GPT-4~\cite{gilardi2023chatgpt}.
After comparing the predicted labels by GPT-4 with the ground truth labels, we find that GPT-4 has a reasonable performance in terms of precision, recall, and F1 score of $0.861, 0.859$, and $0.857$, respectively. 
Such results confirm the superior capability of GPT-4 in our annotation task and we then use it to label all responses, resulting in $23,920$ misinformation-believing, $20,296$ misinformation-disbelieving, and $11,333$ neutral responses out of $55,549$ responses.
The distribution of the response count per counter-reply is shown in Figure~\ref{fig:response_count}.

\section{User Response Characterization}\label{sec:response_chacracterization}

\subsection{Taxonomy of User Response}

Different users respond differently to misinformation correction messages~\cite{wang2022factors, zhang2022investigation}. 
Analyzing these fine-grained behavioral differences in social correction benefits understanding the impact of social correction, so as to promote the social correction that has corrective effects and demote the ones that have backfire effects. 
To this end, we first taxonomize user responses to social correction. 
Given that responses believing or disbelieving in misinformation primarily serve as crucial signals to indicate user reactions, we omit the remaining neutral responses -- responses that neither believe nor disbelieve in misinformation -- due to their minimal effects.
After following similar works on reply analysis~\cite{buchanan2022reading}, we exhaustively analyze manually annotated responses and finally create the taxonomy of user responses in Table~\ref{tab:taxonomy_of_user_response}, presenting user actions employed to demonstrate the corresponding response type. We also assign each user response to its salient action type on a random sample of 220 and finally compute the ratio of each action to the number of total actions within the same response type. 
\begin{table*}[t!]
    \centering
    \setlength\extrarowheight{-5pt}
    \begin{tabular}{p{0.11\linewidth} | p{0.35\linewidth} | p{0.4\linewidth} | p{0.05\linewidth}}
     \toprule
       Response type & User actions employed to demonstrate the corresponding response type & Examples & Ratio \\
        \bottomrule 
        \toprule
      \multirow{4}{=}{Misinformation-disbelieving response} 
      & Endorse those who counter-reply
      & ``You are right'' 
      & 7.692\% \\
      \cline{2-4}
      & Confirm the counter-misinformation 
      & ``I checked the information as well and it is correct'' 
      & 15.385\% \\
      \cline{2-4}
      & Debunk or counter the misinformation again
      & ``Again, the first tweet is misinformative and a bait!'' 
      & 23.077\% \\
      \cline{2-4}
      & Provide additional evidence or supporting information to back up the counter-misinformation 
      & ``Additionally, COVID-19 vaccine only generates spike protein in the cell to protect our body'' 
      & 53.846\% \\
      \bottomrule
      \toprule
      \multirow{5}{=}{Misinformation-believing response} 
      & Refute or insult those who counter-reply 
      & ``You are completely wrong'' or ``You are such a fool to think in this way'' 
      & 19.791\% \\
      \cline{2-4}
      & Reject the counter-reply 
      & ``What you said does not make sense to me. The reasoning in your reply is faulty.'' 
      & 7.291\% \\
      \cline{2-4}
      & Repeat, rephrase, or reconfirm the original misinformation tweet 
      & ``No. The vaccine actually is the gene therapy. It aims to change our DNA.'' and ``The first tweet about the vaccine is correct'' 
      & 19.792\% \\
      \cline{2-4}
      & Provide additional ``evidence'', anecdotal experience, or supporting information to back up the misinformation 
      & ``I knew my grandfather took the vaccine and died later. So, please do not take it'' 
      & 50.000\% \\
      \cline{2-4}
      & Add new types of related misinformation 
      & ``Besides changing our DNA, the vaccine is actually developed by Bill Gates to depopulate the people. Take caution!'' 
      & 3.125\% \\
      \bottomrule
    \end{tabular}
    \caption{Taxonomy of user responses based on employed user actions within each response type. 
    }
    \label{tab:taxonomy_of_user_response}
\end{table*} 
As we can see in Table~\ref{tab:taxonomy_of_user_response}, there are more kinds of user actions in misinformation-believing responses than in misinformation-disbelieving responses. The potential reason is that when people are backfired by social correction, they will explore various ways to express their anger toward the counter-replies. 
Regarding the ratio of different user actions, we notice that providing additional information to back up misinformation or counter-misinformation is the primary action. This phenomenon may be explained that when users debate or discuss conflicting things on the Internet, they are more likely to include ``evidence'' to convince others. 
Altogether, users act differently toward social correction, yet primarily by adding ``supporting information'' to justify their beliefs.

\subsection{Types of Counter-reply That Gets User Responses}\label{sec:counter_reply_group}

\begin{figure}[!htbp]
    \centering
    \includegraphics[width=0.8\columnwidth]{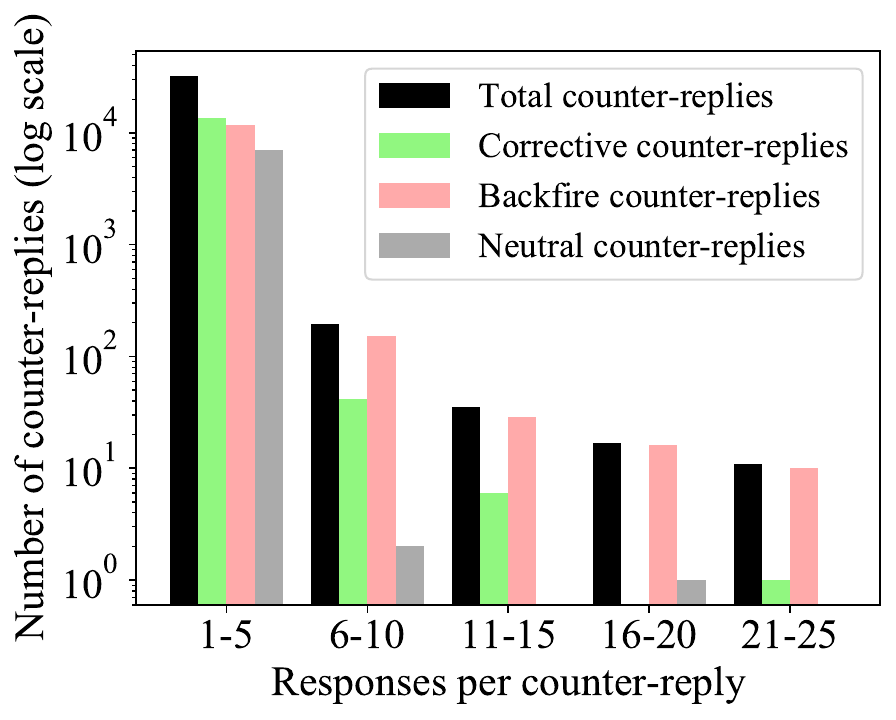}
    \caption{Distributions of the total number of counter-replies (black), number of corrective counter-replies (green), number of backfire counter-replies (red), and number of neutral replies (gray) based on response per counter-reply, each presented on a log scale. 
    }
    \label{fig:reply_count}
\end{figure}

Different counter-replies can lead to different responses ( i.e., responses showing belief, disbelief, or neither belief nor disbelief in misinformation). 
To investigate these counter-replies, 
we group them into four categories based on the response information and then compare them across these four categories. 
Practically, we first categorize replies based on the number of misinformation-believing, misinformation-disbelieving, and neutral responses a counter-reply has. We follow similar research works~\cite{ma2023characterizing, miyazaki2023fake} and neglect replies that have more than 25 responses since they only account for $0.218\%$ of all counter-replies and these ``super-replies'' may skew the analysis~\cite{ma2023characterizing}.  
Finally, we have the following categories for counter-replies:
\begin{itemize}
    \item Corrective counter-reply: Counter-replies contain at least one misinformation-disbelieving response but no \\ misinformation-believing responses.
    \item Backfire counter-reply: Similarly, counter-replies contain at least one misinformation-believing response but no \\ misinformation-disbelieving responses.
    \item Neutral counter-reply: Counter-replies that only contain neutral responses.
    \item Dual counter-reply: Counter-replies contain both \\ misinformation-believing and misinformation-disbelieving responses.
\end{itemize} 
Finally, we identify $13,482$ corrective replies, $11,893$ backfire replies, $7,005$ neutral replies, and $2,385$ dual replies in our dataset. 
Due to the lowest volume of dual replies, we follow the similar research works regarding dual labels~\cite{he2021racism} and do not emphasize them, which can be worth exploring in future studies.
The distribution of the reply count regarding responses per counter-reply is shown in Figure~\ref{fig:reply_count}.

\subsection{Analysis of Counter-reply}\label{sec:feature_analysis}

Analyzing and comparing counter-replies contributes to identifying salient features that are correlated with corrective or backfire effects.
Given the prominent impacts of corrective and backfire counter-replies on users~\cite{chan2017debunking, wittenberg2020misinformation}, we focus on these two kinds of counter-replies for the comparison analysis. 
Particularly, we build on related tweet analysis works~\cite{ma2023characterizing} and analyze the linguistic, engagement, and poster attribute features~\cite{ma2023characterizing} as well as the counter-misinformation property~\cite{he2023reinforcement} features of replies, as follows: 
\begin{enumerate}
    \item \textbf{Reply linguistic attributes}, to analyze the degree to which the reply falls into meaningful personal, psychological, topical, emotional, and other content-related categories.
    \item \textbf{Reply engagement attributes}, to analyze how much the reply interacts with online users.
    \item \textbf{Reply poster attributes}, to analyze the behavior, popularity, and status of the user behind the counter-reply.
    \item \textbf{Counter-misinformation property attributes}, to analyze the extent to which the reply demonstrates the desirable property required for successful debunking backed up by the communication theory~\cite{he2023reinforcement}. 
\end{enumerate}
\begin{table*}[!htbp]
    \centering
    \setlength\extrarowheight{-5pt}
    \begin{tabular}{p{0.15\linewidth} | p{0.8\linewidth}}
     \toprule
      Attribute type & List of attributes  \\
        \midrule
      \multirow{3}{=}{Reply linguistic} & \tabitem Number of words in the reply. \\
      & \tabitem VADER~\cite{hutto2014vader} positive sentiment, negative sentiment, and compound sentiment of the reply. \\
      & \tabitem For each of the 65 dimensions of the LIWC~\cite{pennebaker2001linguistic} 2007 lexicon, the number of words for that dimension. \\
      \midrule
      \multirow{1}{=}{Reply engagement} & \tabitem Number of replies, likes, retweets, and quote of a reply. \\
      \midrule
      \multirow{3}{=}{Reply poster} & \tabitem Number of followers, and number of users following. \\
      & \tabitem Whether the replier is verified (1) or not (0). \\
      & \tabitem Total number of tweets the replier has posted since account creation. \\
      \midrule
      \multirow{3}{=}{Counter-misinformation property} & \tabitem Politeness score of the reply, computed as the total number of politeness-related linguistic strategy instances in the reply as proposed by~\cite{danescu2013computational}. \\
      & \tabitem Refutation score of the reply, obtained by the existing off-the-shelf classifier to indicate to what extent the reply is refuting the misinformation tweet~\cite{he2023reinforcement}. \\
      & \tabitem Evidence score of the reply, derived by checking the existence of high-credibility and fact-checking URLs in the reply~\cite{Micallef2020}. \\
      \bottomrule
    \end{tabular}
    \caption{List of linguistic, engagement, poster, and counter-misinformation property attributes for the counter-reply analysis. 
    }
    \label{tab:attr_list}
\end{table*}
Table~\ref{tab:attr_list} displays the full list of attributes we statistically study\footnote{This statistical test was performed using Welch's unequal variances \textit{t}-test between corrective and backfire counter-replies.} within each of these four categories. 

\subsubsection{Linguistic Attribute Analysis}

First, we find that corrective replies are 5.409\% more positive ($p < 0.005$)\footnote{All p-values are calculated using the Welch’s unequal variances t-tests.} and 8.581\% less negative ($p < 0.0001 $) than backfire replies. 
We find similar results for the ``negative emotion'' dimension of the LIWC lexicon ($p < 0.05 $). 
This implies that positive tones of counter-replies convey optimistic attitudes to convince users to believe in counter-misinformation, while negative tones attract more attention and friction, and therefore, have more backfiring. 
Regarding the number of words in the tweet, both corrective and backfire replies have a similar length of text containing around 23 words. No statistical significance is found between these two groups.
After LIWC lexicon analysis~\cite{pennebaker2001linguistic}, we identify that backfire replies contain higher usage of affective language (words and phrases that appeal more to emotions) than corrective replies ($p < 0.05$). This indicates that those who continue to believe in misinformation when encountering counter-misinformation posts tend to gravitate more towards replying to counter-replies that induce a stronger emotional reaction. 
Some research works find a similar role of emotional content affecting users' resistance to correction~\cite{ecker2022psychological}. 
Additionally, corrective replies mention more words related to family while backfire replies say more death-related emotions ($p < 0.05 $).

\subsubsection{Engagement Attribute Analysis}
In this section, we examine the impact of engagement attributes on whether counter-replies have corrective or backfire effects. We compare the number of total likes, retweets, quotes, and replies that counter-replies receive.
Because these engagements serve different purposes and have different functionalities on the platform, it is worth analyzing these metrics separately. 
Particularly, we find that corrective replies have more retweets ($0.875$ vs. $0.565$ Avg. retweets per reply; $p < 0.001 $) and likes ($8.629$ vs. $6.218$ Avg. likes per reply; $p < 0.001 $) but fewer replies ($1.357$ vs. $1.753$ Avg replies per reply; $p < 0.001 $) than backfire replies while they share a similar number of quotes ($0.064$ vs $0.065$ Avg. quotes per reply) with no statistical difference. 
These findings may imply that the endorsement through more retweets and likes increases the believability of counter-replies~\cite{morris2012tweeting}, thus having corrective effects. %
In turn, we can also interpret that the misinformation-disbelieving responses make the corrective counter-reply more convincing, finally having more likes and retweets~\cite{boyd2010tweet}. This mutually-reinforced effect demonstrates the importance of engagement attributes in the analysis.

\subsubsection{Poster Attribute Analysis}
We first examine the impact of the user being verified on the counter-reply having corrective or backfire effects. 
We find that the proportion of accounts sending corrective counter-replies that are verified is higher than those sending backfire counter-replies ($0.021$ vs. $0.009$ the proportion of verified accounts, $p < 0.001$). 
Once the account is verified, the audience will be more likely to think it is credible and believe in the counter-misinformation, demonstrating corrective effects.
Likewise, unverified accounts may decrease the credibility of the counter-reply, having backfire effects.  
Similar findings are also identified on another poster feature -- the total number of tweets since account creation. Particularly, on average, those having corrective counter-replies have more total tweets since account creation than those having backfire counter-replies ($p < 0.001 $). 
The potential explanation can be that more tweets indicate more active and engaged repliers, thus enhancing their credibility and having corrective effects. Fewer or no tweets make the audience question the validity of the accounts. 
Regarding the number of followers and followings, even though we do not find a statistical difference in followers, interestingly, we find that those having corrective counter-replies have more followings ($p < 0.001$).

\subsubsection{Counter-misinformation Property Analysis}
Considering the context of counter-misinformation in our analysis, we also examine the three 
properties that have been shown to be crucial in effective counter-misinformation messages~\cite{he2023reinforcement, chan2017debunking}: politeness, evidence, and refutation.

Following the existing work~\cite{danescu2013computational, ma2023characterizing}, we compute the politeness score of each reply and then compare the average politeness scores between the two groups. Our results find that corrective replies are 4.678\% more polite than backfire replies ($p < 0.01$). This result is consistent with the existing theory that polite debunking works better than impolite debunking~\cite{he2023reinforcement, chan2017debunking}.
Regarding evidence, we check the existence of high-credibility and fact-checking URLs in counter-replies, as suggested by ~\citet{Micallef2020}. The result shows that the proportion of counter-replies that have highly credible or fact-checking URLs is 8.137\% higher in corrective replies than in backfire replies. The reason may be that these URLs increase the believability of the counter-reply, finally having corrective effects.

Interestingly, we notice that results in refutation scores are different from the existing theory.
Particularly, the refutation score reveals the degree to which the reply refutes the misinformation tweet. 
The higher the score is, the more explicitly the reply refutes the misinformation tweet, which is needed for effective countering~\cite{chan2017debunking}.
Note that, the refutation score - where we measure the relationship between the misinformation tweet and counter-reply - is not the same as the previously examined negative sentiment. 
In practice, after computing the refutation score of each reply using the existing classifier~\cite{he2023reinforcement} and comparing the average scores between the two categories, we find that corrective replies have lower refutation scores than backfire replies ($p < 0.0001$, and Cohen's $d = 0.202$\footnote{Cohen's d here refers to the unweighted
Cohen's d values.}). 
Even if higher refutation scores are expected in corrective replies~\cite{chan2017debunking}, our result is still explainable considering when we add more refutation statements in replies, the emotions of some audience can be triggered~\cite{bayer2010reading}.
This implies that when countering misinformation in real-world scenarios, we need to attend to the degree of refutation to which we reject the false information and avoid the backfire simultaneously.

\section{User Response Prediction}

In this section, our primary objective is to address the research question: "Given a counter-reply, can we predict whether it will have a corrective, backfire, or neutral effect", as described in Section~\ref{sec:task_objective}.

Being able to accurately predict future interactions following a counter-reply, we can identify sets of online misinformation posts where the counter-reply is organically working, as well as those requiring additional countermeasures. 
Finally, we can pinpoint instances of the counter-replies that do not work such that the associated misinformation tweets can be proactively and carefully countered by other users to curb the spread of misinformation.

\subsection{Dataset}
To answer the above research question, we use the aforementioned dataset in Section~\ref{sec:response_chacracterization}. 
Particularly, we divide the counter-replies into three sets: 
(1) corrective counter-replies; 
(2) backfire counter-replies;
and (3) neutral counter-replies, as defined in Section~\ref{sec:counter_reply_group}. The sizes of the three sets are $13,482$, $11,893$, and $7,005$. 

\subsection{Experiment Setup}
After choosing the dataset, we follow similar approaches in tweet prediction tasks~\cite{Micallef2020, zahera2019fine} 
by building a multi-class classifier.
We utilize the set of attributes described in Section~\ref{sec:feature_analysis} as features.
As shown in the existing tweet prediction work~\cite{Micallef2020}, the semantic information from textual embedding benefits the prediction task. Thus, we also generate the embedding vector for each reply using RoBERTa~\cite{liu2019roberta}. 
Finally, we concatenate the above feature vectors to form a reply feature vector to comprehensively represent the reply and use it for classification.

\subsection{Classifier Creation and Evaluation} 
Following similar tweet or general text classification tasks~\cite{Micallef2020, he2021petgen}, 
we deploy widely-used machine learning classifiers including Logistic Regression, XGBoost, and a Feed-forward Neural Network containing a single hidden layer, using the feature vector as input. 
During the experiment, 10-fold cross-validation is deployed, and we report precision, recall, and F-1 score as the performance metrics.

\begin{table}[!t]
    \centering
    \begin{tabular}{lrrr}
    \hline    
    Method &  Precision &  Recall &  F1 score \\ \hline
    Logistic Regression         &  0.753   & 0.787   &  0.769    \\
    XGBoost                     &   0.814  &  0.764  & 0.788   \\
    Neural Network              &  0.832   &  0.801  &  0.816 \\ \hline
    \end{tabular}
    \caption{Classification performance of whether a counter-reply will have a corrective, backfire, or natural effect.}
    
    \label{tab:result_rq}
\end{table}

The classification result is shown in Table~\ref{tab:result_rq}. As we can see, the model performance is reasonably acceptable. 
Especially, the neural network achieves the best performance regarding precision, recall, and F-1 score; this finding is also found in other similar tweet classification tasks~\cite{Micallef2020}.
This high performance offers the ability to effectively predict whether a counter-reply will have a corrective, backfire, or neutral effect, enabling fact-checkers and social media platforms to organically prioritize counter-replies identified as more likely to backfire.

\section{Discussion and Conclusion}

In this paper, we curate a large-scale conversation-style dataset containing user responses to social correction and build a taxonomy to present different types of these user responses.
We also study the text- and user-level properties of counter-replies that have corrective or backfire effects. The in-depth analysis shows that counter-replies expressing positive sentiments and politeness and having evidence are more likely to have corrective effects. 
Our result also shows that counter-replies that have corrective effects have a higher amount of retweet and like engagement that expresses endorsement. 
Moreover, we develop a well-performed classifier to predict whether a counter-reply will have a corrective, backfire, or neutral effect.
In sum, our work comprehensively demonstrates that the user response to social corrections has implications regarding what kinds of social corrections work better, and sheds light on how to combat misinformation by social correction.

There are still some limitations in our work. 
First, we only utilize user responses to determine the impact of counter-reply, which is acceptable because user responses can provide both positive and negative feedback through the expressed disbelief and belief in misinformation respectively. 
However, the number of responses is usually small for one counter-reply, and the signals from the user engagement (e.g., retweets and likes) are not considered together to form a comprehensive evaluation of counter-replies in our work.
Another notable limitation is its exclusive focus on Twitter. The dynamics of post engagement and information exchange can vary significantly across different online platforms~\cite{micallef2022cross}, potentially influencing the nature of social correction and user response to it. 
Besides, even if GPT-4 demonstrates commendable performance in annotation tasks~\cite{mao2023gpteval}, our study utilizes it exclusively for annotating responses, rather than extending its use to tweets and replies, which does not perform a uniform annotation process for all data points. The reason is that the employment of GPT-4 is constrained by its high API costs. In contrast, traditional low-cost BERT-based classifiers for tweets and replies yield satisfactory results, with both F1 scores exceeding 0.8, which aligns well with the requirements of our research. 
We also admit that our results will depend on the reliability of classifiers to accurately identify misinformation tweets and counter-replies for downstream analysis.
Additionally, our study is limited to English language text since we filter out misinformation tweets in other languages. The dynamics in other languages could present different patterns in misinformation spread and correction.
Furthermore, our analysis is confined to discussions around COVID-19 vaccines, a topic that has garnered widespread attention due to the global impact of the COVID-19 pandemic. This focus may not fully represent the dynamics of other 
prevalent misinformation topics~\cite{treen2020online}, such as climate change misinformation, where the specific countering text and demographics of posters could influence interaction patterns in distinct ways. 
Finally, we only examine the text information while other modalities like images and videos can manifest various patterns and actions in social correction.

For future work, we can first consider combining user engagements (e.g., likes and retweets of counter-replies) with user responses together to comprehensively determine the impact of counter-replies. 
Second, we could extend our analysis to the user networks of misinformation posters, those who counter-reply, and responders to the counter-replies to investigate the potential phenomenon of networked ``echo chamber''~\cite{del2015echo}.
This would involve examining the followers and followees of these users, as well as the prevalence of misinformation and counter-misinformation within these networks, to identify network attributes that might influence the effect of counter-replies:  backfire or corrective effects. 
In addition, accurately predicting whether a counter-reply can have a corrective, backfire, or neutral effect opens up opportunities for field studies to investigate how specific characteristics of counter-replies might affect a user's belief in misinformation.

\textbf{{ACKNOWLEDGMENTS}} 
This research/material is based upon work supported by NSF grant CNS-2154118. Any opinions, findings and conclusions or recommendations expressed in this material are those of the author(s) and do not necessarily reflect the position or policy of NSF and no official endorsement should be inferred.


\bibliographystyle{ACM-Reference-Format}
\balance
\bibliography{main}

\appendix

\end{document}